\documentclass[aps,prb,showpacs,floats]{revtex4}
\usepackage{graphicx,psfig}
\usepackage{dcolumn}
\usepackage{bm}

\usepackage{amssymb}
\usepackage{epsfig}

\begin{document}
\title{Energy landscape properties studied by symbolic sequences}
\author{Rolf Schilling}
\affiliation{Institut f\"ur Physik, Johannes Gutenberg Universit\"at Mainz, \\
Staudinger Weg 7, D-55099 Mainz, Germany}
\date{\today}
\medskip
Dedicated to Serge Aubry on the occasion of his 60th birthday
\begin{abstract}
We investigate a classical lattice system with $N$ particles. The
potential energy $V$ of the scalar displacements is chosen as a
$\phi ^4$ on-site potential plus interactions. Its stationary
points are solutions of a coupled set of nonlinear equations.
Starting with Aubry's anti-continuum limit it is easy to establish
a one-to-one correspondence between the stationary points of $V$
and symbolic sequences $\bm{\sigma} = (\sigma_1,\ldots,\sigma_N)$
with $\sigma_n=+,0,-$. We prove that this correspondence remains
valid for interactions with a coupling constant $\epsilon$ below a
critical value $\epsilon _c$ and that it allows the use of a
''thermodynamic'' formalism to calculate statistical properties of
the so-called ``energy landscape'' of $V$. This offers an
explanation why topological quantities of $V$ may become singular,
like in phase transitions. Particularly, we find the saddle index
distribution is maximum at a saddle index $n_s^{max}=1/3$ for all
$\epsilon < \epsilon_c$. Furthermore there exists an interval
($v^*,v_{max}$) in which the saddle index $n_s$ as function of
average energy $\bar{v}$ is analytical in $\bar{v}$ and it
vanishes at $v^*$, above the ground state energy $v_{gs}$, whereas
the average saddle index $\bar{n}_s$ as function of energy $v$ is
highly nontrivial. It can exhibit a singularity at a critical
energy $v_c$ and it vanishes at $v_{gs}$, only. Close to
$v_{gs},\; \bar{n}_s(v)$ exhibits power law behavior which even
holds for noninteracting particles.
\end{abstract}

\maketitle

\section{Introduction}

\label{1} We consider a classical N-particle system in three
dimensions with potential energy $V(\vec{x}_1,\ldots,\vec{x}_N)$.
Both the Newtonian dynamics and the thermodynamical behavior is
obtained from the knowledge of the function $V$. Therefore one may
ask: What are the characteristic features of $V$ which are crucial
for the dynamics and thermodynamics of the N-particle system? From
a mathematical point of view the stationary points, in case of
non-degeneracy also called critical or Morse points, yield
important information on $V$. These points are solutions of the
set of coupled, \textit{nonlinear} equations:
\begin{equation} \label{eq1}
\frac{\partial V}{\partial \vec{x}_n} (\vec{x}_1,\ldots,
\vec{x}_N)=0\quad, \; n = 1, \ldots, N \quad .
\end{equation}

 These solutions are denoted by
 ${\bf{x}}^{(\alpha)}=(\vec{x}^{(\alpha)}_1,\ldots,\vec{x}_N^{(\alpha)}), \;
 \alpha =1,2,\ldots,M_N$, where $M_N$ is believed to be
 exponentially in $N$. Here some comments are in order. First, if
 the system is homogeneous and isotropic any translation, rotation
 and reflection of $\{\vec{x}_n^{(\alpha)}\}$ is a stationary point
 as well, with the same potential energy. Accordingly there is a
 continuous degeneracy of $\{\vec{x}_n^{(\alpha)}\}$. Choosing the
 variables $\vec{x}_n$, e.g.~with respect to the center of mass
 and fixing the orientation one gets rid of this degeneracy. This
 is assumed in the following. Second, if $V$ is
 {\textit{harmonic}}, i.e.
 \begin{equation}\label{eq2}
 V(\vec{x}_1,\ldots, \vec{x}_N) = \frac 1 2 \sum \limits _{{n,m} \atop{=1}} ^N
 \sum \limits ^3 _{{i,j}\atop{=1}} M^{ij}_{nm} x^i_n x^j_m
 \end{equation}
 Eq.~(\ref{eq1}) becomes a set of {\textit{linear}} equations:
 \begin{equation}\label{eq3}
 \sum^N_{m=1} \sum \limits_{j=1} ^3 M^{ij}_{nm} x^j_m =0\;, \quad
 n= 1,\ldots,N ;\; i = 1,2,3
 \end{equation}
 In the generic case where $\det(M^{ij}_{nm})\neq 0$, there is one
 solution, $\vec{x}_n \equiv 0$, only. Consequently, a
 {\textit{necessary}} condition for exponentially many solutions
 is the {\textit{nonlinearity}} of Eq.~(\ref{eq1}).
 Having found all stationary points they can be characterized by
 their saddle index
 $n_s({\bf{x}}^{(\alpha)})=N_s({\bf{x}}^{(\alpha)})/(3N).\quad
 N_s({\bf{x}}^{(\alpha)})$ is the number of unstable directions,
 i.e. the number of negative eigenvalues of the Hessian
 $((\partial ^2V/\partial x_n^i \partial
 x_m^j)({\bf{x}}^{(\alpha)})$. Therefore $n_s$ varies between zero
 and one.

 The role of $n_s$ for the dynamics is obvious. If the trajectory
 ${\bf{x}}(t)$ in configuration space is mostly close to stationary
 points with $n_s$ very close to zero, i.e. ${\bf{x}}(t)$ is close
 to local minima of $V(\vec{x}_1,\ldots,\vec{x}_N)$, then the
 motion will mainly be due to thermal activation, i.e.
 hopping-like. In contrast, if ${\bf{x}}(t)$ is mostly close to
 stationary points with $n_s$ almost equal to one, a
 diffusive-like dynamics may occur. Recent MD-simulations for
 liquids have determined the average saddle index $\bar{n}_s$ as
 function of temperature $T$ \cite{1} and as function of the
 potential energy $v$ per particle \cite{2,3}. It has been found that
 $\bar{n}_s$ decreases with decreasing temperature and energy and its
 extrapolation vanishes at a temperature $T^*$ and an energy $v^*$,
 respectively. It is interesting that $T^*$ is very close to the
 mode coupling glass transition temperature $T_c$, at which a
 transition from ergodic to nonergodic dynamics takes place. This
 finding corresponds to the change from diffusive- to hopping-like
 motion, when decreasing the saddle index from one to zero.

 There is evidence that the saddle index also plays a role for
 equilibrium phase transitions. For smooth, finite range and
 confining potentials $V$ it has been proven \cite{4} that a
 \textit{necessary} condition for an equilibrium phase transition
 is a singular change of the topology of the manifold
 \begin{equation}\label{eq4}
 {\mathcal{M}}_N(v)=\{(\vec{x}_1,\ldots,\vec{x}_N)|V(\vec{x}_1,\ldots,\vec{x}_N)\leq
 v\}.
 \end{equation}
 Particularly, if $M_N(v,n_s)$ is the number of stationary points
 of ${\mathcal{M}}_N(v)$ with saddle index $n_s$, the Euler
 characteristic
 \begin{equation}\label{eq5}
 \chi(v)= \sum \limits_{N_s=0}^{3N} (-1)^{N_s} M_N(v,N_s/3N)
 \end{equation}
 can be singular at a critical energy $v_c$. However, for nonconfining
 potentials the unattainability of a purely topological criterion
 for the existence of equilibrium phase transition has been
 claimed \cite{5}. Besides these interesting results it has been found for
 a two-dimensional $\phi^4$-model from numerical computations
 \cite{6}, for the exactly solvable mean-field $XY$ model \cite{7}
 and mean-field k-trigonometric model \cite{8} that $v_c$
 coincides with $N^{-1}\langle V \rangle (T_c)$, the
 {\textit{canonical}}
 average of the potential energy per particle at the phase transition point
 $T_c$. However, for a mean-field $\phi^4$-model it has been shown that
 $v_c \neq N^{-1} \langle V \rangle (T_c)$ \cite{9,10}.

 Finally we want to mention an interesting result for the saddle
 index distribution function $P_N(n_s)$. For a binary
 Lennard-Jones system with up to $N=13$ particles there is strong
 evidence that $P_N(n_s)$ is Gaussian with a maximum at $n_s^{max}
 \approx 1/3$ \cite{3}. However, the relevance of this result for
 dynamics and thermodynamics is not yet clear.

 This exposition so far has demonstrated that the potential energy
 surface (PES) characterized by stationary points and their saddle
 indices is of physical importance. For further applications of
 the energy landscape description the reader may consult the
 textbook by D. J. Wales \cite{11}.

 It is the motivation of the present paper to derive a
 relationship between ${\bf{x}}^{(\alpha)}$ and symbolic sequences
 $\bm{\sigma}=(\sigma_1,\ldots,\sigma_N)$ where $\sigma_n$ takes
 values from an ``alphabet''. This will be explained in the next
 section. Particularly the usefulness of this relationship for the
 calculation of statistical properties of the PES will be shown.
 To make this relation more explicit we will investigate in the
 3.~section noninteracting particles in an on-site potential. In
 the 4.~section it will be explored how the features of the PES
 change under switching on an interaction between the particles.
 The final section contains a discussion of the results and some
 conclusions.


\section{Description by symbolic sequences}
\label{2} 
\begin{figure}[t]
\includegraphics[width=8cm,angle=-90]{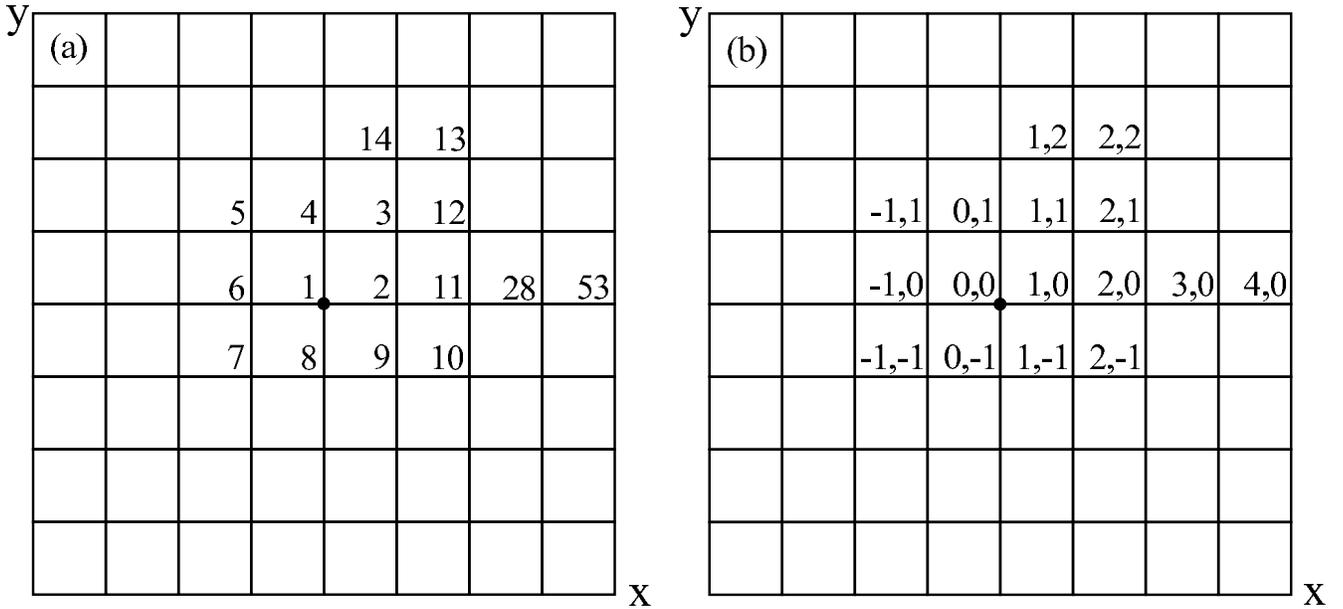}
\caption{\label{fig1} Labelling of the square lattice by (a)
natural numbers and (b) by ``sequences'' $(\sigma_1,\sigma_2)$
with integer values of $\sigma_n$.}
\end{figure}

For some nonlinear dynamical systems it has been proven
(see e.g.~Refs. \cite{12,13}) that there is a one-to-one
correspondence between the orbits and symbolic sequences.
Therefore, the brilliant observation by Aubry \cite{14} that
Eq.~(\ref{eq1}) for the one-dimensional Frenkel-Kontorova model is
identical to the standard map, already provides a link between
stationary points and symbolic sequences. This link exists for a
certain class of one-dimensional models \cite{15}. But we think
that this relationship may be more general. Let us choose a simple
potential with two degrees of freedom:
\begin{equation}\label{eq6}
V(x_1,x_2)= \sin x_1 \; \sin x_2
\end{equation}
Its stationary points form a square lattice with lattice constant
$\pi/2$. The question arises how to label these stationary points.
The first guess is to count the points as shown in
Fig.~\ref{fig1}a.

However, there is a drawback since two neighboring points do not
possess ``neighboring'' labels (see e.g.~ the points with labels
28 and 53). In order that the labelling preserves the local
arrangement of the points one has to choose an ``alphabet'', which
are the integer numbers $\sigma _n \in \mathbb{Z}$. Then the
``sequences'' ($\sigma_1,\sigma_2)$ provide a labelling for which
the local properties are preserved (cf.~Figure \ref{fig1}a+b).
Similarly, the counting of the stationary points of an
\textit{arbitrary} PES by $\alpha = 1,2,\ldots,M_N$, as done in
the 1.~section, is not compatible with the local properties of
these points. Although the explicit determination of an
``alphabet'' ${\mathcal{A}}$ and of the one-to-one correspondence
between the stationary points of an arbitrary PES and symbolic
sequences ($\sigma_1,\ldots,\sigma_N$) with $\sigma_n \in
{\mathcal{A}}$ in general is not possible we believe, however,
that such a relationship may exist for certain potential functions
$V$ . This will be proven in the third and fourth section for a
certain class of functions $V(\vec{x}_1,\ldots,\vec{x}_N)$. Of
course, for a finite system there will be a finite number of
stationary points which can be labelled by $\alpha =
1,2,3,\ldots$. But similarly to the simple model (\ref{eq6}) this
will be not the appropriate labelling.

Thus, let us assume that we have found ${\mathcal{A}}$ and the
one-to-one mapping:
\begin{equation}\label{eq7}
\bm{\sigma}=(\sigma_1,\ldots,\sigma_N) \rightarrowtail
(\vec{x}_1(\bm{\sigma}),\ldots,\vec{x}_N(\bm{\sigma}))={\bf{x}}(\bm{\sigma})
\end{equation}
between the symbolic sequences $\bm{\sigma}$ and the stationary
points $\bm{x}^{(\alpha)} \hat{=} \bm{x}(\bm{\sigma})$. Their
potential energy is given by:
\begin{equation}\label{eq8}
E(\bm{\sigma}) = V(\bf{x}(\bm{\sigma})).
\end{equation}
Note that depending on the specific potential energy $V$ the
sequence may contain more than $N$ symbols, e.g.
$\bm{\sigma}=(\sigma _1, \ldots,\sigma_{3N})$. The Hessian at
these stationary points is a function of $\bm{\sigma}$, too.
Therefore its number of negative eigenvalues, $N_s$, and
accordingly the saddle index $n_s$ is a function of $\bm{\sigma}$,
only:
\begin{equation}\label{eq9}
n_s(\bm{\sigma}) = N_s (\bm{\sigma})/(3N).
\end{equation}

In the remainder of this section we will demonstrate the
usefulness of the mapping (\ref{eq7}) for the calculation of some
properties of the PES. One of the most interesting quantities is
the joint probability density $P_N(v,n_s)$ of stationary points
with potential energy $v$ per particle and saddle index $n_s$.
Without making use of relation (\ref{eq7}) it can be represented
as follows:
\begin{eqnarray}\label{eq10}
P_N(v,n_s)= M^{-1}_N \int \prod \limits_{n=1}^N d^3x_n |\det
(\frac{\partial ^2 V(\bf{x})} {\partial x ^i_n \partial _{x_m}^j})
| (\prod \limits _{n=1}^N \prod \limits _{i=1}^3 \delta
(\frac{\partial V({\bf{x}})} {\partial x ^i_n}))\\ \nonumber
\delta (v-N^{-1}V({\bf{x}})) \delta (n_s- N^{-1}N_s({\bf{x}})).
\end{eqnarray}
One of the main technical problems for the evaluation of
$P_N(v,n_s)$ from Eq.~(\ref{eq10}) is the occurrence of the
{\textit{modulus}} of the determinant of the Hessian. This modulus
prevents the use of an integral representation by Grassmann
variables. For mean-field spin glass models it has been shown that
the modulus can be neglected, at least for low energies \cite{16}.
Now, making use of Eqs.~(\ref{eq7})-(\ref{eq9}), this probability
density can also be represented by:
\begin{equation}\label{eq11}
P_N(v,n_s)=N^2M_N^{-1} \sum \limits _{\bm{\sigma}}
\delta(Nv-E(\bm{\sigma}))\delta (Nn_s-N_s(\bm{\sigma}))
\end{equation}
which is rewritten as follows:
\begin{equation}\label{eq12}
P_N(v,n_s)= \frac{N^2M^{-1}_N}{(2 \pi)^2} \int \limits ^\infty _{-
\infty} d \lambda \int \limits ^\infty _{- \infty}d \mu
\;e^{iN[\lambda v + \mu n _s-f(\lambda,\mu)]}
\end{equation}
with the ``free energy'' per particle
\begin{equation}\label{eq13}
f(\lambda,\mu)=iN^{-1} \ln Z(\lambda,\mu ; N)
\end{equation}
and the ``canonical partition function''
\begin{equation}\label{eq14}
Z(\lambda, \mu ; N)= \sum \limits_{\bm{\sigma}} \exp [-i(\lambda E
(\bm{\sigma}) + \mu N_s(\bm{\sigma}))].
\end{equation}
This demonstrates that, e.g.~$P_N(v,n_s)$ can be obtained from a
``canonical ensemble'' with probability density:
\begin{equation}\label{eq15}
\rho (\bm{\sigma}) = \frac 1 Z e^{-i(\lambda E (\bm{\sigma}) + \mu
N_s (\bm{\sigma}))} \;.
\end{equation}
$\lambda$ can be interpreted as an inverse ``temperature'' and
$\mu$ as a ``field'' acting as a bias on the number of negative
eigenvalues of the Hessian.

In the thermodynamic limit $N \rightarrow \infty$ the saddle point
solutions $\lambda ^*(v,n_s), \\ \mu^* (v,n_s)$ of
\begin{equation}\label{eq16}
v = \frac{\partial f}{\partial \lambda} (\lambda, \mu),\quad n_s =
\frac{\partial f}{\partial \mu} (\lambda, \mu)
\end{equation}
yield up to a normalization constant:
\begin{equation}\label{eq17}
P_N(v,n_s) \sim e^{+Ns(v,n_s)}
\end{equation}
with the configurational entropy (per particle) of stationary
points with energy per particle $v$ and saddle index $n_s$:
\begin{equation}\label{eq18}
s(v,n_s)=i\{\lambda ^*(v,n_s) v+ \mu
^*(v,n_s)n_s-f(\lambda^*(v,n_s),\mu ^*(v,n_s))\}.
\end{equation}
$s(v,n_s)$ allows the calculation of the saddle index as function
of energy. However, there are two possibilities. First, we
determine the maximum $\bar{v}$ of $s(v,n_s)$ for $n_s$ fixed.
This yields the {\textit{saddle index}} $n_s(\bar{v})$ as function
of the {\textit{average energy}}. Second, keeping the energy $v$
fixed the maximum of $s(v,n_s)$ yields the {\textit{average saddle
index}} $\bar{n}_s(v)$ as function of {\textit{energy}}. These two
functions, $n_s$ and $\bar{n}_s$ are not identical, in general.
From (\ref{eq18}) we find:
\begin{equation}\label{eq19}
\lambda ^*(\bar{v},n_s(\bar{v}))=0, \quad \mu ^*
(v,\bar{n}_s(v))=0
\end{equation}
where Eq.~(\ref{eq16}) has been applied.

The saddle index distribution $P_N(n_s)= \int \limits ^\infty _{-
\infty} dv P_N(v,n_s)$ can be represented as:
\begin{equation}\label{eq20}
P_N(n_s)=\frac{NM_N^{-1}}{2 \pi} \int \limits ^\infty _ {- \infty}
d \mu e^{iN[\mu n_s-f_0(\mu)]}
\end{equation}
with
\begin{equation}\label{eq21}
f_0(\mu)=f(0,\mu).
\end{equation}
For $N \rightarrow \infty$ the saddle point solution $\mu
^*_0(n_s)$ of
\begin{equation}\label{eq22}
n_s = \frac {df_0}{\partial \mu} (\mu)
\end{equation}
leads to
\begin{equation}\label{eq23}
P_N(n_s) \sim e^{Ns_0(n_s)}
\end{equation}
with
\begin{equation}\label{eq24}
s_0(n_s)=i\{\mu^*_0(n_s)n_s - f_0(\mu_0^*(n_s))\}.
\end{equation}
Its maximum position $n_s^{max}$ is given by:
\begin{equation}\label{eq25}
\mu ^*_0(n_s^{max})=0 \;.
\end{equation}

This last paragraph has demonstrated the usefulness of the
``thermodynamic'' formalism based on a canonical ensemble
(\ref{eq15}) in the space of symbolic sequences. This fact also
allows to understand why there can be a {\textit{singular}}
topological change, e.g.~as a function of $v$. There is a lower
critical dimension $d_{low}$ such that there is a critical
``temperature'' $\lambda _c^{-1}$ for $d >d_{low}$ at which
$f(\lambda,\mu)$ is singular. This singularity then also occurs in
the ``Legendre transform'' $s(v,n_s)$ of $f(\lambda,\mu)$.

In the third and fourth section we will show that this
``thermodynamic'' formalism applied to a certain class of lattice
models allows an analytical calculation of $P_N(v,n_s)$ from which
the saddle index distribution and the energy dependence of the
saddle index can be derived.

\section{Noninteracting particles}
\label{3}

We consider a crystal with $N$ lattice sites. Let $x_n$ be the
scalar displacement of particle $n$ from its lattice site with
lattice vector $\vec{R}_n$. Then we describe the potential energy
by a $\phi^4$-model:
\begin{equation}\label{eq26}
V(\bm{x})= \sum \limits _{n=1} ^N V_0(x_n)+\epsilon V_1(\bm{x}),\;
\quad \epsilon >0
\end{equation}
with the double-well-like on-site potential:
\begin{equation}\label{eq27}
V_0(x)=- hx - \frac 1 2 x^2 + \frac 1 4 x^4\;,\quad h>0
\end{equation}
and interaction $V_1(\bm{x})$. Choosing an appropriate energy and
length scale any general quartic on-site potential can be put into
the form of Eq.~(\ref{eq27}).

Inspired by Aubry's anti-continuum limit (originally called
anti-integrable limit) \cite{17,18} we start with $\epsilon =0$,
i.e. neglecting the interaction. Then Eq.~(\ref{eq1}) becomes:
\begin{equation}\label{eq28}
x^3_n - x_n = h\;.
\end{equation}
If $h < h_c=2/(3\sqrt{3})$ there are three different real roots
$x_n=x_{\sigma _n}(h),\; \sigma _n=+,0,-$. Therefore the
stationary points are given by
\begin{equation}\label{eq29}
{\bf{x}}_0(\bm{\sigma}) = (x_{\sigma_1},\ldots,x_{\sigma_N})\;,
\end{equation}
i.e.~we have found the one-to-one correspondence between
stationary points and symbolic sequences with an ``alphabet''
${\mathcal{A}}={+,0,-}$. Using Eqs.~(\ref{eq8}),(\ref{eq27}) and
(\ref{eq29}) we find for the energy of the stationary points
\begin{equation}\label{eq30}
E_0(\bm{\sigma}) = \sum \limits _{n=1}^N [e_0+e_1 \sigma _n+e_2
\sigma_n^2]
\end{equation}
with
\begin{eqnarray}\label{eq31}
e_0=v_0,\; e_1 &=& \frac 1 2 (v_+-v_-)\;,\; e_2 = -v_0 + \frac 1 2
(v_++v_-)\\ \nonumber v_\sigma &=& V_0(x_\sigma).
\end{eqnarray}
These results are obvious, since the stationary points of $V$ for
$\epsilon =0$ are uniquely determined by the stationary points of
the local potential $V_0(x)$. The number of extrema of $V_0(x)$
determine the ``alphabet'' ${\mathcal{A}}$ and their energy
$V_0(x_\sigma)$ yields the coefficients $e_\nu, \nu =0,1,2$ of
$E_0(\bm{\sigma})$. Furthermore the function $N_s(\bm{\sigma})$ is
also easily determined. Since we identify the maximum of
$V_0(x_n)$ with $\sigma_n=0$, its absolute and local minimum,
respectively, with $\sigma_n=+$ and $\sigma _n =-$, it is:
\begin{equation}\label{eq32}
N_s(\bm{\sigma}) = \sum \limits _{n=1}^N(1-\sigma _n^2)\,.
\end{equation}

Having found the ``alphabet'' ${\mathcal{A}}$, the mapping
(\ref{eq7}) as well as $E_0(\bm{\sigma})$ and $N_s(\bm{\sigma})$
we can calculate $P_N(v,n_s)$ as described in the previous
section. The calculation of the ``partition function''
(\ref{eq14}) is easy. One gets from Eq.~(\ref{eq13}):
\begin{equation}\label{eq33}
f(\lambda,\mu)=(\lambda e_0 + \mu)+ i \ln [1+2e^{-i(\lambda
e_2-\mu)} \cos \lambda e_1]\;.
\end{equation}
The saddle point solutions of Eq~(\ref{eq16}) are easily
determined leading to $P_N(v,n_s)$ from Eq.~(\ref{eq17}) with:
\begin{equation}\label{eq34}
s(v,n_s)=-[n_-(v,n_s)\ln n_-(v,n_s)+n_+(v,n_s)\ln
n_+(v,n_s)+n_s\ln n_s]
\end{equation}
where:
\begin{equation}\label{eq35}
n_\pm(v,n_s)= \pm(v_--v_+)^{-1}[(v_\mp -v)+(v_0-v_\mp)n_s].
\end{equation}
for {\textit{asymmetric}} double well, i.e.~$v_+ \neq v_-$. Quite
analogously one finds the saddle point $\mu ^*_0 (n_s)$:
\begin{equation}\label{eq36}
\mu^*_0(n_s)=i \ln \frac{1-n_s}{2n_s}
\end{equation}
leading to:
\begin{equation}\label{eq37}
s_0(n_s)=-[n_s \ln n_s+(1-n_s) \ln (1-n_s)/2]
\end{equation}
for symmetric and asymmetric double well potentials.

To determine the relation between the saddle index and energy for
{\textit{symmetric}} double well, i.e.~for $h=0$, we have to
realize that $n_s$ and $v$ are no longer independent variables.
This can easily be seen from Eq.~(\ref{eq30}), taking $e_1=0$ into
account, which follows from Eq.~(\ref{eq31}). Then we arrive at
\begin{equation}\label{eq38}
E(\bm{\sigma})=N v_++(v_0-v_+)N_s(\bm{\sigma})
\end{equation}
where Eqs.~(\ref{eq31}) and (\ref{eq32}) were used. This yields
immediately:
\begin{equation}\label{eq39}
v(n_s)=v_++(v_0-v_+)n_s\;,
\end{equation}
and no distinction between $n_s(\bar{v})$ and $\bar{n}_s(v)$
exists. Because of the dependence of both variables $v$ and $n_s$
on each other their joint probability density reduces to
$P_N(n_s)$ which is proportional to the probability density of
$v$. From Eqs.~(\ref{eq25}) and (\ref{eq36}) we find for symmetric
and asymmetric double wells:
\begin{equation}\label{eq40}
n_s^{max}=1/3
\end{equation}
which is the maximum of $s_0(n_s)$ (Eq.~(\ref{eq37})) and
therefore the maximum of the saddle index distribution $P_N(n_s)$.
$P_N(n_s)$ is a Gaussian for $|n_s-n_s^{max}|=O(1/\sqrt{N})$, in
agreement with the numerical result for Lennard-Jones clusters
\cite{3}.

\begin{figure}[t]
\includegraphics[width=8cm,angle=-90]{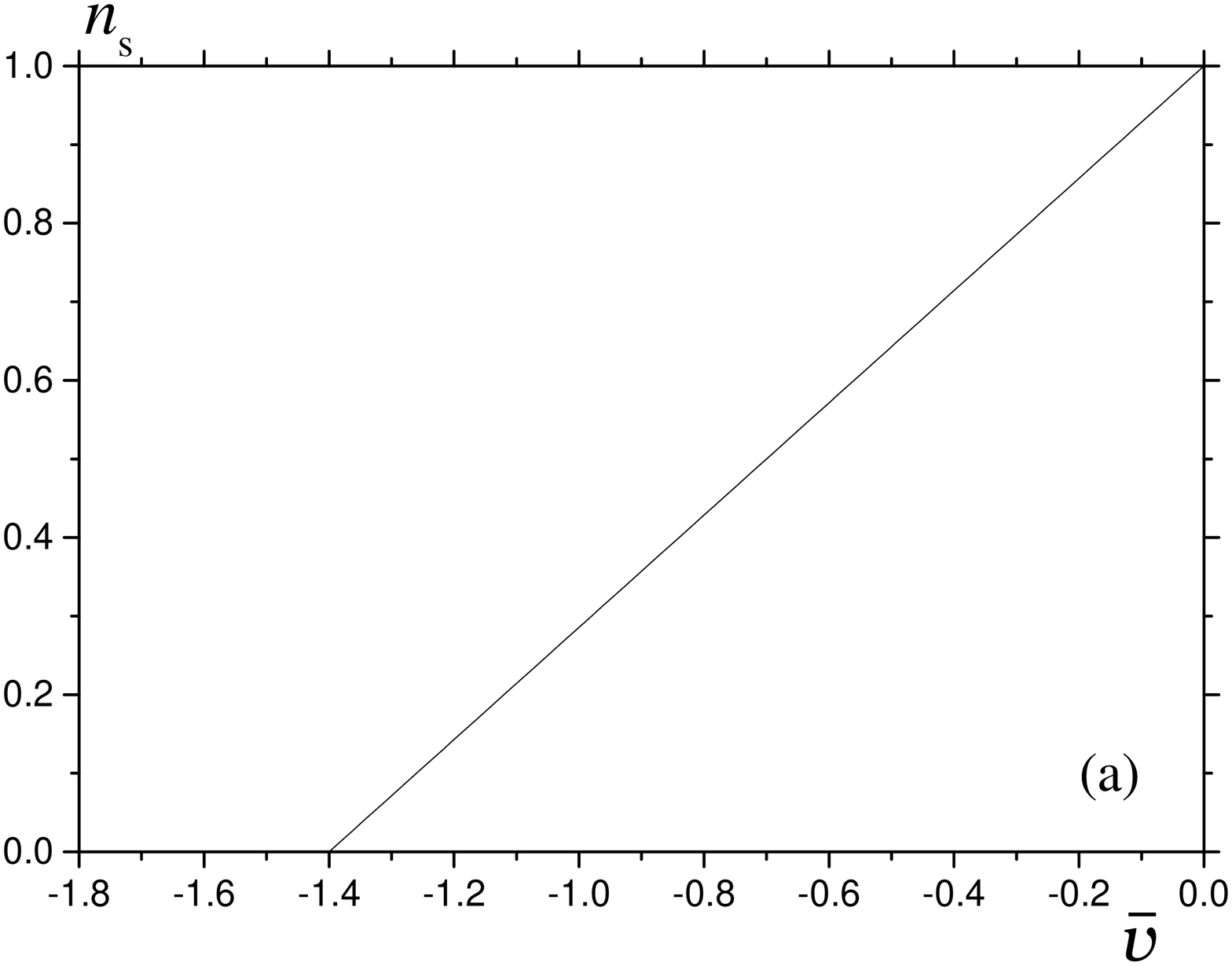}
%
\includegraphics[width=8cm,angle=-90]{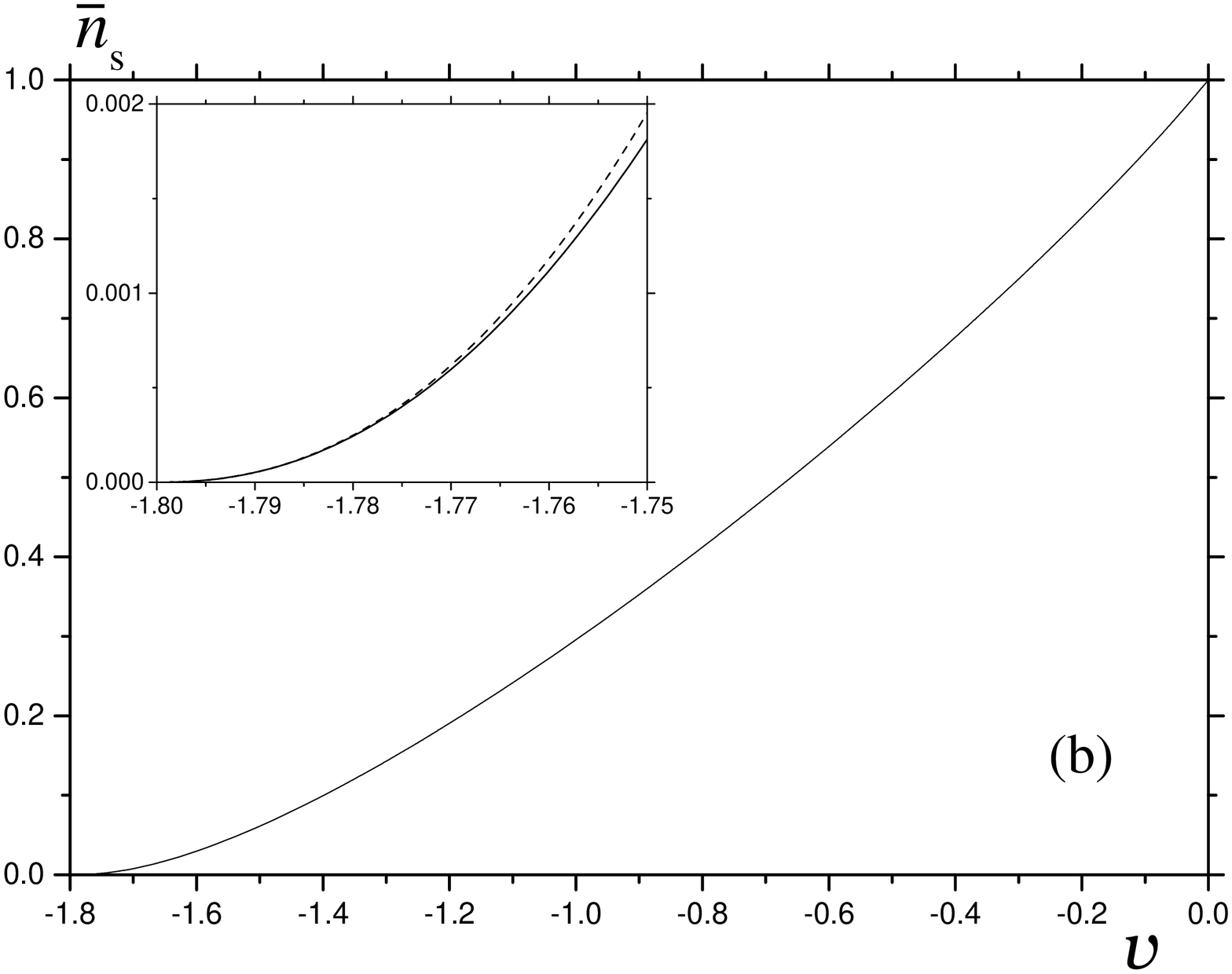}
\caption{\label{fig2} (a) saddle index $n_s$ as function of the
average energy $\bar{v}$, (b) average saddle index $\bar{n}_s$ as
function of the energy $v$. Both figures are for noninteracting
particles $(\epsilon =0)$ with $v_0=0$, $v_-=-1$ and $v_+=-1.8$.
The inset in Fig.~\ref{fig2}b shows the comparison of the exact
result for $\bar{n}_s(v)$ from Eq.~(\ref{eq42})(solid line) with
the leading power law from Eq.~(\ref{eq43}) (dashed line).}
\end{figure}

Let us return to the {\textit{asymmetric}} double well. The
functions $n_s(\bar{v})$ and $\bar{n}_s(v)$ can be obtained from
the maximum of $s(v,n_s)$ for fixed $n_s$ and fixed $v$,
respectively. As a result we get
\begin{equation}\label{eq41}
n_s(\bar{v})= \frac{(v_-+v_+)-2\bar{v}}{-2v_0+(v_-+v_+)}
\end{equation}
and
\begin{equation}\label{eq42}
\Big[ \frac{n_+(v,\bar{n}_s)}{\bar{n}_s}\Big ]
^{\frac{v_0-v_-}{v_--v_+}} = \Big[
\frac{n_-(v,\bar{n}_s)}{\bar{n}_s} \Big ]
^{\frac{v_0-v_+}{v_--v_+}}
\end{equation}
with $v_-, v_0,v_+$ from Eq.~(\ref{eq31}) and $n_\pm
(v,\bar{n}_s)$ from Eq.~(\ref{eq35}). $n_s$ is a linear function
of the average energy $\bar{v}$. It vanishes at $\bar{v}^*= \frac
1 2 (v_-+v_+)$ which is above the ground state energy $v_{gs} =
v_+$, since we assumed asymmetric double wells. $n_s$ becomes one
for $\bar{v}=v_0$, the height of the unstable extremum of
$V_0(x)$. $n_s(\bar{v})$ is presented in Figure \ref{fig2}a.
Eq.~(\ref{eq42}) is an implicit one for $\bar{n}_s(v)$, which can
not be solved analytically. The numerical result is shown in
Figure \ref{fig2}b. There are two features to be mentioned. First,
$\bar{n}_s(v)$ vanishes at the ground state energy $v_{gs}=v_+$,
only. Second, it can be proven analytically that it exhibits a
power law behavior in $v$ close to $v_{gs}$ (cf.~the inset of
Fig.~\ref{fig2}b):
\begin{equation}\label{eq43}
\bar{n}_s(v) \simeq (\frac{v-v_{gs}}{v_--v_{gs}})^{\delta _0} \;
+O ((\frac {v-v_{gs}}{v_--v_{gs}})^{\delta _0+1})+O ((\frac
{v-v_{gs}}{v_--v_{gs}})^{2\delta _0-1})
\end{equation}
for $\frac {v}{v_{gs}} - 1 \ll 1$ with an exponent:
\begin{equation}\label{eq44}
\delta _0= \frac {v_0-v_+}{v_--v_+} > 1 \;.
\end{equation}

It is interesting that we find for $\bar{n}_s(v)$  (in contrast to
$n_s(\bar{v})$) a nonanalytical $v$-dependence close to $v_{gs}$,
despite the neglection of interactions. In the next section we
will show that this power law also exists in case of interactions.

So far we have demonstrated that the ``thermodynamic'' formalism
can be applied to get the saddle index distribution,
$n_s(\bar{v})$ and $\bar{n}_s(v)$. $n_s(\bar{v})$ and
$\bar{n}_s(v)$ also follow directly from the saddle points
$\lambda^*$ and $\mu ^*$ (cf.~Eq.~\ref{eq19}). This has not been
used, because $\lambda ^*(v,n_s)$ is a rather lengthy expression.
The results (\ref{eq37}), (\ref{eq40}) concerning the saddle index
distribution, and (\ref{eq41}) for $n_s(\bar{v})$ can be obtained
much easier as follows. The number of stationary points
characterized by $N_\sigma$, which is the number of $\sigma _n$ of
$\bm{\sigma}$ equal to $\sigma \in \{-,0,+\}$, is given by
\begin{equation}\label{eq45}
\frac{N!}{N_+!N_0!N_-!}
\end{equation}
which for fixed saddle index $n_s=N_0/N$ has its maximum weight
for $N_+=N_-= \frac 1 2(N-N_s)$. Substituting this into expression
(\ref{eq45}) we get by use of Stirling's formula:
\begin{equation}\label{eq46}
P_N(n_s) \sim e^{Ns_0(n_s)}
\end{equation}
with $s_0(n_s)$ from Eq.~(\ref{eq37}). The maximum position
$n_s^{max}$ is also obvious. If $N_+^{max}= N_0^{max}\equiv
N_-^{max}\equiv N/3$, expression (\ref{eq45}) takes its maximum
value. Therefore $n_s^{max}=N_0^{max}/N=1/3$. Next we use
$E(\bm{\sigma})$ from Eq.~(\ref{eq30}) to calculate
$\bar{v}(n_s)=N^{-1} \langle E(\bm{\sigma})\rangle$. Using again
that the maximum weight of (\ref{eq45}), for fixed $n_s$, follows
for $N_+=N_-$, we get:
\begin{equation}\label{eq47}
\langle \sum \limits _n \sigma _n \rangle = N_+ - N_- =0\;, \quad
\langle \sum \limits _n \sigma _n^2 \rangle = N - N_0
\end{equation}
Substituting this result into Eq.~(\ref{eq30}) yields:
\begin{equation}\label{eq48}
\bar{v}(n_s)= \frac 1 2 (v_- + v_+) +[v_0- \frac 1 2 (v_-+v_+)]
n_s
\end{equation}
from which Eq.~(\ref{eq41}) is reproduced. The question arises:
Which of these features remain valid in the presence of
interactions? This will be discussed in the next section.

\section{Interacting particles}
\label{4}

Now we will study the influence of interactions on the results
derived in the previous section. $V(\bm{x})$ is given by
Eq.~(\ref{eq26}) and (\ref{eq27}) where we will assume that the
interaction energy $V_1(\bm{x})$ is (i) at least twice
differentiable and (ii) does not grow faster than quartic. To
investigate the labelling of all stationary points of $V(\bm{x})$
for $\epsilon > 0$ we adopt the method used by MacKay and Aubry
and MacKay and Sepulchre to prove the existence of breathers
\cite{19} and to investigate multistability in networks \cite{20},
respectively, i.e.~the implicit function theorem. In the last
section we have proven for $\epsilon =0$ that all stationary
points $\bm{x}_0(\bm{\sigma})\equiv \bm{x}(\epsilon
=0,\bm{\sigma})$ are uniquely labelled by symbolic sequences
$\bm{\sigma},\; \sigma _n=+,0,-$, provided $h<h_c$. The
eigenvalues $\lambda_\nu (0,\bm{\sigma})$ of the Hessian
$(\epsilon =0)$ are nonzero for all $\bm{\sigma}$ and $h<h_c$
with:
\begin{equation}\label{eq49}
\textrm{sign} \; \lambda _{\nu =n} (0,\bm{\sigma}) =-1+2\sigma
^2_n
\end{equation}
This implies that
\begin{equation}\label{eq50}
\det (\frac {\partial ^2 V}{\partial x_n\partial x_m}
(\bm{x}_0(\bm{\sigma}))) \neq 0
\end{equation}
for all $\bm{\sigma}$. Let us introduce the functions
\begin{equation}\label{eq51}
\phi _n(\epsilon,\bm{x})=\frac {\partial V}{\partial x_n}
(\epsilon , \bm{x})\;,
\end{equation}
which are at least one times differentiable. Note that we have
made the $\epsilon$-dependence of $V$ explicit. Then
Eq.~(\ref{eq1}) reads:
\begin{equation}\label{eq52}
\phi _n(\epsilon, \bm{x})=0\;, \quad n = 1,\ldots, N\quad .
\end{equation}

We have found solutions $\bm{x}_0(\bm{\sigma})$ for $\epsilon =0$:
\begin{equation}\label{eq53}
\phi _n (0,\bm{x}_0(\bm{\sigma}))=0,\; n=1,\ldots,N\quad .
\end{equation}
Because the Jacobian:
\begin{equation}\label{eq54}
\frac{\partial(\phi_1,\ldots,\phi_N)}{\partial (x_1,\ldots,x_N)}
(\bm{x}_0(\bm{\sigma}))= \det (\frac{\partial ^2V}{\partial x_n
\partial x_m}(\bm{x}_0(\bm{\sigma}))) \neq 0
\end{equation}
for \textit{all} $\bm{\sigma}$ and $0 \leq h <h_c$, we can apply
the implicit function theorem \cite{21} which guarantees the
existence of a neighborhood $\tilde{U}(\bm{\sigma})$ of $(\epsilon
=0, \bm{x}_0 (\bm{\sigma})) \in \mathbb{R}^{N+1}$, of an open set
$W(\bm{\sigma})=[0,\epsilon _c (\bm{\sigma})]\subset \mathbb{R}$
and of functions $x_n(\epsilon, \bm{\sigma}), n= 1,\ldots,N$ (at
least one times differentiable) such that
\begin{equation}\label{eq55}
\phi _n (\epsilon, \bm{x}(\epsilon, \bm{\sigma}))=0\;,
\end{equation}
for all $\epsilon \in W(\bm{\sigma})$ and all $\bm{\sigma}$. This
leads to a nonvanishing critical value
\begin{equation}\label{eq56}
\epsilon _c = \min \limits _{\bm{\sigma}}
\epsilon_c(\bm{\sigma})>0
\end{equation}
such that $\bm{x}(\epsilon,\bm{\sigma})$ are stationary points of
$V(\bm{x})$ for $0 \leq \epsilon <\epsilon_c$. Therefore the
one-to-one correspondence between these points and symbolic
sequences $\bm{\sigma}$ with $\sigma _n \in \{+,0,-\}$ holds for
$0 \leq \epsilon <\epsilon _c$ and $h<h_c$.

At $\epsilon_c$ a bifurcation occurs at which stationary points
disappear and new ones may be created. Accordingly the
``alphabet'' and/or the mapping between stationary points and
symbolic sequences will change at $\epsilon_c$. This bifurcation
is signalled by the vanishing of at least one of the eigenvalues
$\lambda _\nu (\epsilon _c, \bm{\sigma})$, for one or more
sequences $\bm{\sigma}$. Consequently, it is
\begin{equation}\label{eq57}
{\textrm{sign}} \lambda _\nu(\epsilon, \bm{\sigma}) =
{\textrm{sign}} \lambda _\nu(0,\bm{\sigma})
\end{equation}
for $0 \leq \epsilon <\epsilon_c$, since a change of sign occurs
at $\epsilon _c$, only. This has the strong implication that the
saddle index of $\bm{x}(\epsilon,\bm{\sigma})$ is identical to
that of $\bm{x}(0,\bm{\sigma})$, which is given by the number
$N_0$ of $\sigma _n=0$. This result can also be put into other
words: Although the relative heights of the stationary points will
change with $\epsilon$, the topology will remain unchanged. By
this we mean that a stationary point with $N_s$ unstable
directions remains a stationary point with $N_s$ unstable
directions for $0 \leq \epsilon <\epsilon _c$. We note, that this
result requires condition (ii) for the function $V_1$. In case
that $V_1(\bm{x})$ grows faster than quartic, additional
stationary points may occur already for arbitrary small values of
$\epsilon$. Validity of (ii) guarantees that
$\bm{x}(\epsilon,\bm{\sigma})$, which are continuously connected
to $\bm{x}(0,\bm{\sigma})$, are the only solutions of
Eq.~(\ref{eq1}).

Some general conclusions can be drawn from this result. Since the
one-to-one correspondence between the stationary points and
$\bm{\sigma}$ is preserved for $0 \leq \epsilon < \epsilon_c$, the
saddle index distribution $P_N(n_s)$ and accordingly
$n_s^{max}=1/3$ remains the same. Although the calculation of
$n_s(\bar{v})$ and $\bar{n}_s(v)$ needs the knowledge of the
energy $E(\bm{\sigma})$ of the stationary points, one can prove
some general properties of both functions without using the
explicit form of $E(\bm{\sigma})$. Let us begin with
$n_s(\bar{v})$ which is obtained from $\bar{v}(n_s)$. As already
shown in the last section $\bar{v}(n_s)$ is directly obtained from
\begin{equation}\label{eq58}
\bar{v}(n_s)=(N M_N(n_s))^{-1} \sum \limits _{\bm{\sigma} \atop
{N_s(\bm{\sigma})=n_sN}} E(\bm{\sigma})
\end{equation}
where $M_N(n_s)$ is the number of $\bm{\sigma}$ with
$N_s(\bm{\sigma})=n_sN$. $n_s=0$ yields \textit{all} local minima
of $V(\bm{x})$, including the ground state. Most of these minima
have an energy above the ground state energy $Nv_{gs}$, i.e.~it
is:
\begin{equation}\label{eq59}
\bar{v}(0) >(NM_N)^{-1}Nv_{gs} M_N(n_s) = v_{gs}\quad .
\end{equation}
For $n_s=1$, i.e.~$N=N_s$, there is only one stationary point, the
maximum of $V(\bm{x})$, with energy $E(0,\ldots,0)$ such that
\begin{equation}\label{eq60}
\bar{v}(1)= N^{-1}E(0,\ldots,0) \equiv v_{max}
\end{equation}
From this we find that $n_s(\bar{v})$ vanishes at
$\bar{v}^*>v_{gs}$ and becomes one at $\bar{v}=v_{max}$. Whether
or not $n_s(\bar{v})$ is always monotonically increasing with
increasing $\bar{v}$ is not clear. In contrast to $n_s(\bar{v})$,
the average saddle index $\bar{n}_s(v)$ is nonzero for all $v >
v_{gs}$. Without presenting a rigorous prove, let us explain why
this should be true. The ground state belongs to $n_s =0$ and is
characterized by $\bm{\sigma}$ with $\sigma ^{gs}_n = + $ or $-$.
This also holds, if it is degenerate. Now, let us choose $K$
particles $j_1,\ldots,j_K$ for which we change $\sigma^{gs}_{j_k}
\in \{+,-\}$ into $\sigma _{j_k}=0$. The corresponding stationary
point has $n_s=K/N$. If $K=1$, we generate a ``defect'' with
excitation energy $\epsilon _{j_1}$. Then $K$ ``defects'' have an
energy $E_{{j_1} \ldots j_K} = N v_{gs}+ \sum \limits _{k=1}^K
\epsilon_{j_k} + $ (...), where (...) is the interaction energy of
the defects. It is obvious that $E_{{j_1}\ldots j_K}/N \rightarrow
v_{gs}$ for $K \rightarrow \infty$, $N \rightarrow \infty$ with
$K/N\rightarrow 0$. ``Defect'' configurations with fixed energy
$v= E_{{j_1}\ldots j_K}/N$ will have fluctuating $K$ with $0 <
K/N$. If we choose $v-v_{gs}$ arbitrary small but nonzero the
average saddle index $\bar{n}_s =\bar{K}/N$ will be small, but
finite, too. Consequently $\bar{n}_s$ can only vanish at $v_{gs}$.
Since there is one stationary point (maximum) with energy
$v_{max}$, only, it must be $\bar{n}_s(v_{max})=1$, i.e.
\begin{equation}\label{eq61}
n_s(\bar{v}=v_{max}) = \bar{n}_s(v_{max})=1.
\end{equation}
Again, it is not obvious whether $\bar{n}_s(v)$ is monotonously
increasing with increasing $v$.

To get more quantitative results for $n_s(\bar{v})$ and
$\bar{n}_s(v)$, we have to specify $E(\bm{\sigma})$. Since the
solutions $\bm{x}(\epsilon, \bm{\sigma})$ are not known exactly
this can only be done approximately. In the following we will
present the crucial steps leaving out technical details. Our
purpose is to derive the qualitative structure of
$E(\bm{\sigma})$. To be explicit we choose harmonic interactions:
\begin{equation}\label{eq62}
V_1(\bm{x})= \frac 1 2 \sum \limits _{n,m} V_{nm}x_n x_m \; ,
\quad V_{nn}\equiv 0.
\end{equation}
with coupling coefficients $V_{nm} = V(\vec{R}_n-\vec{R_m})$.
Using $V_1(\bm{x})$ from Eq.~(\ref{eq62}), Eq.~(\ref{eq1}) takes
the form:
\begin{equation}\label{eq63}
-x_n + x^3_n - h =-\epsilon \sum \limits _m V_{nn}x_m
\end{equation}
which can be solved by iterations:
\begin{equation}\label{eq64}
-x_n^{(1)}+(x_n^{(1)})^3-h=-\epsilon \sum \limits _m
V_{nm}x_m^{(0)}\;, \quad x_m^{(0)}=x_m(\epsilon
=0,h)=x_{\sigma_m}(h)
\end{equation}
etc. We remind the reader that $x_{\sigma_n}(h)$ are the roots of
Eq.~(\ref{eq28}). If $|h-\epsilon \sum _m V_{nm}x_{\sigma
_m}(h)|<h_c$ for all $\bm{\sigma}$, Eq.~(\ref{eq64}) has three
real roots:
\begin{equation}\label{eq65}
x_n^{(1)}(\epsilon ,\bm{\sigma})=x_{\sigma_n}(h-\epsilon \sum
\limits _m V_{nm} x_{\sigma _m}(h))
\end{equation}
which will be expanded with respect to $\epsilon$:
\begin{eqnarray}\label{eq66}
x_n^{(1)}(\epsilon, \bm{\sigma})&=& x_{\sigma _n}(h) - \epsilon
x_{\sigma _n}' (h) \sum \limits _m V_{nm} x_{\sigma _m}(h) +
\\ \nonumber &+& \frac 1 2 \epsilon ^2x_{\sigma _n}^{''} (h) \sum \limits _{m,m'}
V_{nm} V_{nm'} x_{\sigma _m}(h)x_{\sigma _{m'}} (h) .
\end{eqnarray}
$\frac{d^\ell x_\sigma (h)}{dh^\ell}$ can be represented as
follows:
\begin{equation}\label{eq67}
\frac{d^\ell x_\sigma (h)}{d h^\ell} = \sum \limits _{i=0}^2 x_i
^{(\ell)}(h)\sigma ^i
\end{equation}
where $x_i^{(\ell)}$ is easily be expressed by the l-th derivative
of $x_\sigma(h),\sigma = +,0,-$. Using Eq.~(\ref{eq67}) we get
from Eq.~(\ref{eq66})
\begin{eqnarray}\label{eq68}
x_n^{(1)} (\epsilon, \bm{\sigma}) &=& \sum \limits _{i_n=0}^2
x_{i_n} (h)\sigma _n^{i_n} + \epsilon \sum \limits _{i_ni_m=0}^2
x_{n,i_n i_m}(h)\sigma_n^{i_n} \sigma _m ^{i_m} + \\ \nonumber &+&
\frac 1 2 \epsilon ^2 \sum \limits ^2 _{i_ni_mi_{m'}=0} x_{n, i_n
i_mi_{m'}}(h) \sigma _n ^{i_n} \sigma _m ^{i_m}
\sigma_{m'}^{i_{m'}} + \cdots
\end{eqnarray}
with:
\begin{equation}\label{eq69}
x_{i_n}(h) \equiv x_{i_n}^{(0)} (h), \;
x_{n,i_ni_m}(h)=-x'_{i_n}(h)\sum \limits _m V_{nm}x_{i_m}(h)\;,
{\textrm{etc.}} \quad .
\end{equation}

The substitution of $x_n^{(1)}(\epsilon, \bm{\sigma})$ from
Eq.~(\ref{eq68}) into $V(\bm{x})$ with $V_1(\bm{x})$ from
Eq.~(\ref{eq62}) leads to:
\begin{eqnarray}\label{eq70}
E^{(1)}(\bm{\sigma})&\equiv & V(\bm{x}^{(1)}(\bm{\sigma})) = \sum
\limits_n [e_0^{(1)}(\epsilon)+e_1^{(1)}(\epsilon) \sigma _n +e
_2^{(1)} (\epsilon) \sigma_n^2] + \\ \nonumber &+& \epsilon \sum
\limits _{n_1 \neq n_2} [A^{(1)}_{n_1n_2} \sigma_{n_1}
\sigma_{n_2} + B^{(1)}_{n_1n_2}(\sigma^2_{n_1} \sigma_{n_2}+
\sigma _{n_1}\sigma_{n_2}^2) + C^{(1)}_{n_1n_2}\sigma _{n_1}^2
\sigma_{n_2}^2]+ \\ \nonumber &+& \epsilon ^2 \sum \limits
_{n_1\neq n_2 \neq n_3 \neq n_1}[A^{(1)}_{n_1n_2n_3} \sigma _{n_1}
\sigma _{n_2} \sigma_{n_3} + B^{(1)}_{n_1n_2n_3}(\sigma ^2_{n_1}
\sigma_{n_2} \sigma_{n_3} +\\ \nonumber &+& \sigma _{n_1}
\sigma_{n_2}^2 \sigma_{n_3}+ \sigma _{n_1} \sigma_{n_2}
\sigma_{n_3}^2)+ C^{(1)}_{n_1n_2n_3} (\sigma ^2_{n_1}
\sigma_{n_2}^2 \sigma_{n_3}+ \\ \nonumber &+& \sigma ^2_{n_1}
\sigma_{n_2} \sigma_{n_3}^2+ \sigma _{n_1} \sigma_{n_2}^2
\sigma_{n_3}^2)+ D^{(1)}_{n_1n_2n_3}\sigma_{n_1} ^2 \sigma_{n_2}^2
\sigma_{n_3}^2] + (\cdots)
\end{eqnarray}
where $(\cdots)$ are four-, five- etc.~body interactions which are
of order $\epsilon ^3, \epsilon ^4$ etc. The coefficients in
capital letters in Eq.~(\ref{eq70}) can be expressed by
$x_{i_n}(h),\; x_{n,i_ni_m}(h)$, etc., $h$ and $V_{nm}$.
$E^{(1)}(\bm{\sigma})$ is a kind of generalized
Blume-Emery-Griffiths model. Using higher order iterates the
corresponding energy $E^{(\nu)}(\bm{\sigma})$ will be similar to
Eq.~(\ref{eq70}) with ``renormalized'' coefficients $e_i^{(\nu)},
\; A^{(\nu)}_{n_1n_2}$, etc. On the other hand it is obvious that
any function $f(\bm{\sigma})$ can be represented by a form as
given on the r.h.s.~of Eq.~(\ref{eq70}). Therefore we choose for
$E(\bm{\sigma})$ the r.h.s.~of Eq.~(\ref{eq70}) without the
superscripts.

We begin with the calculation of $n_s(\bar{v})$. Taking
$\lambda^*(\bar{v},n_s(\bar{v}))=0$ (cf.~Eq.~(\ref{eq19})) into
account the saddle point equations (\ref{eq16}) reduce to:
\begin{eqnarray}\label{eq71}
\bar{v}= \frac {1}{N\cdot Z(0,\mu^*;N)} \sum \limits
_{\bm{\sigma}} E(\bm{\sigma})e^{-i\mu^*N_s(\bm{\sigma})}\;, \\
\nonumber  n_s = \frac{1}{N\cdot Z(0,\mu^*;N)} \sum \limits
_{\bm{\sigma}} N_s(\bm{\sigma})e^{-i\mu^*N_s(\bm{\sigma})}
\end{eqnarray}
$\mu^*(n_s)$ is easily obtained from the second equation of
(\ref{eq71}):
\begin{equation}\label{eq72}
\mu ^*(n_s)=i \ln \frac{2n_s}{1-n_s}
\end{equation}
and the first one can be written as follows:
\begin{eqnarray}\label{eq73}
\bar{v} &=& \frac 1 N  \{ \sum \limits _n
[e_0(\epsilon)+e_1(\epsilon)\langle \sigma _n\rangle _0 +
e_2(\epsilon) \langle \sigma _n^2 \rangle _0] + \\ \nonumber &+&
\epsilon \sum \limits _{n_1 \neq n_2} [A_{n_1n_2}\langle \sigma
_{n_1} \rangle _0 \langle \sigma_{n_2} \rangle_0 +
B_{n_1n_2}(\langle \sigma _{n_1}^2 \rangle _0 \langle
\sigma_{n_2}\rangle_0+ \langle \sigma _{n_1}\rangle _0 \langle \sigma ^2_{n_2}\rangle _0)+ \\
\nonumber &+& C_{n_1n_2}\langle \sigma _{n_1} ^2\rangle _0 \langle
\sigma_{n_2}^2\rangle _0] + \cdots \}
\end{eqnarray}
with
\begin{equation}\label{eq74}
\langle f(\sigma)\rangle _0 = \sum \limits _{\sigma
=+,0,-}f(\sigma)e^{-i\mu ^*(1-\sigma^2)}/\sum \limits _{\sigma =
+,0,-} e^{-i\mu ^*(1-\sigma ^2)}\; .
\end{equation}
Taking $\mu ^*$ from Eq.~(\ref{eq72}) into account one easily
finds:
\begin{equation}\label{eq75}
\langle \sigma _n\rangle _0 =0\quad , \; \langle \sigma_n^2\rangle
_0 = 1-n_s\;.
\end{equation}
This result is obvious, since for $N_s=n_sN$ fixed the maximum
weight of expression (\ref{eq45}) is obtained for
$N_+=N_-=N(1-n_s)/2$ which immediately implies Eq.~(\ref{eq75}).
Introducing the result from Eq.~(\ref{eq75}) into
Eq.~(\ref{eq73})\ yields:
\begin{equation}\label{eq76}
\bar{v}(n_s)= e_0 (\epsilon) + e_2(\epsilon)(1-n_s)+ \epsilon
\tilde{C}_2(0)(1-n_s)^2+ \epsilon ^2 \tilde{C}_3(0)(1-n_s)^3 +
\cdots
\end{equation}
with:
\begin{equation}\label{eq77}
\tilde{C}_2(0)= \sum\limits_{n_2 \neq 0}C_{0n_2}\quad , \quad
\tilde{C}_3(0)= \sum \limits _{0 \neq n_2 \neq n_3 \neq 0}
C_{0n_2n_3},\cdots
\end{equation}
where the lattice translational invariance has been taken into
account. For $\epsilon$ small enough, $\bar{v}(n_s)$ is monotonous
in $n_s$ and we can solve Eq.~(\ref{eq76}) for $n_s(\bar{v})$:
\begin{equation}\label{eq78}
n_s(\bar{v})=1+ \frac{\bar{v}-(e_0(\epsilon)+
e_2(\epsilon))}{e_2(\epsilon)} - \epsilon \frac{\tilde{C}_2(0)}
{e_2(\epsilon)} (\frac{\bar{v}-(e_0(\epsilon)+
e_2(\epsilon))}{e_2(\epsilon)})^2 \\ \nonumber +0(\epsilon^2).
\end{equation}

Note that the bare one-particle quantities $e_i$
(cf.~Eqs.~(\ref{eq30}), (\ref{eq31})) are ``renormalized'' to
$e_i(\epsilon)$ which depend on the coupling constants $V_{nm}$.
Putting in Eq.~(\ref{eq78}) $\epsilon =0$, one recovers (by use of
Eq.~(\ref{eq31})), the result (\ref{eq41})). The interaction
between the particles has \textit{two} effects. First, it
``renormalizes'' the one-particle coefficients $e_i,i=0,1,2$ such
that $e_1(\epsilon)\neq 0$, in general. This corresponds to a
non-symmetric, effective on-site potential. Second, $n_s(\bar{v})$
becomes nonlinear. Its curvature depends on the sign of the
effective coupling constant $\tilde{C}_2(0)$.

The calculation of $\bar{n}_s(v)$ is much more involved, because
one has to perform averages with respect to $\exp[-i \lambda
E(\bm{\sigma})]$, which contains the interactions. This is in
contrast to the averaging with $\exp[-i\mu N_s (\bm{\sigma})]$
(cf.~Eq.~(\ref{eq71})) which factorizes. Therefore an analytically
exact determination of $\bar{n}_s(v)$ for \textit{all} $v$ is not
possible. But, in the following we will demonstrate that the
``thermodynamic'' formalism can be applied to get $\bar{n}_s(v)$
for $v/v_{gs}-1 \ll 1$. In that limit we can perform a
\textit{cumulant expansion}. The simplest case is a
``ferro-elastic'' ground state, i.e.~$\sigma_n^{gs} \equiv +$, as
for the non-interacting particles. For $v$ close to
$v_{gs}=E(+,\ldots,+)/N$ only such stationary points exist for
which a low concentration of $\sigma _n's$ deviate from $+$.
Therefore we introduce ``defect'' variables:
\begin{equation}\label{eq79}
\tau _n=1-\sigma _n \quad .
\end{equation}
The ground state belongs to $\tau _n = 0$. $\tau _n = 1$ or 2
indicates a ``defect''. Replacing in Eq.~(\ref{eq70}) $\sigma_n$
by $\tau _n$ yields:
\begin{equation}\label{eq80}
E(\bm{\tau})=E(+,\ldots,+)+E_0(\bm{\tau})+ E_1(\bm{\tau})
\end{equation}
with
\begin{eqnarray}\label{eq81}
E_0(\bm{\tau}) &=& \sum \limits _n [\tilde{e}_1(\epsilon)\tau_n+
\tilde{e}_2(\epsilon) \tau ^2_n] \\ \nonumber E_1(\bm{\tau})&=&
\sum \limits _{n_1 \neq n_2} [\tilde{A}_{n_1n_2}(\epsilon)\tau
_{n_1}\tau_{n_2}+\tilde{B}_{n_1n_2}(\epsilon)(\tau_{n_1}^2\tau_{n_2}+
\tau_{n_1}\tau_{n_2}^2) +\tilde{C}_{n_1n_2}(\epsilon)
\tau_{n_1}^2\tau_{n_2}^2]+\\ \nonumber &+& \ldots
\end{eqnarray}
where $\tilde{A}_{n_1n_2},\tilde{B}_{n_1n_2}$, etc. contain
different orders in $\epsilon$ and vanish for $\epsilon =0$. They
can easily be expressed by $A_{n_1n_2}, B_{n_1n_2}$, etc. For
$\tilde{e}_1(\epsilon),\tilde{e}(\epsilon)$ one obtains:
\begin{eqnarray}\label{eq82}
\tilde{e}_1(\epsilon)+\tilde{e}_2(\epsilon)=
e_0(\epsilon)-v_{gs}(\epsilon) \equiv
v_0(\epsilon)-v_{gs}(\epsilon) \\ \nonumber
2(\tilde{e}_1(\epsilon)+2\tilde{e}_2(\epsilon)) =
e_0(\epsilon)-e_1(\epsilon)+e_2(\epsilon)-v_{gs}(\epsilon) =
v_-(\epsilon)-v_{gs}(\epsilon)
\end{eqnarray}
where $v_0(\epsilon)$ and $v_-(\epsilon)$ are obtained from
Eq.~(\ref{eq31}) by replacing $e_i$ by $e_i(\epsilon)$. Using a
cumulant expansion we get for the ``free energy'':
\begin{eqnarray}\label{eq83}
f(\lambda,\mu)=v_{gs}\lambda + f_0(\lambda, \mu)+ \lambda \frac 1
N \langle E_1(\bm{\tau})\rangle _0 + \\ \nonumber + i \lambda ^2
\frac {1}{2N} [\langle (E_1(\bm{\tau}))^2\rangle _0- (\langle
E_1(\bm{\tau})\rangle )^2] + \cdots
\end{eqnarray}
The quantities with subscript 0 are obtained with the unperturbed
``canonical ensemble'':
\begin{equation}\label{eq84}
\rho_0(\bm{\tau}) = \frac {1}{Z_0} e^{-i(\lambda E_0(\bm{\tau})+
\mu N_s(\bm{\tau}))}
\end{equation}
where $N_s(\bm{\tau})$ is obtained from Eq.~(\ref{eq32}) by use of
Eq.~(\ref{eq79}). Then the saddle point equations (\ref{eq16}) are
as follows:
\begin{eqnarray}\label{eq85}
v-v_{gs}&=& \frac{\partial f_0}{\partial \lambda} + \frac 1 N
(\langle E_1(\bm{\tau})\rangle _0+ \lambda \frac
{\partial}{\partial \lambda} \langle E_1(\bm{\tau})\rangle_0) +
\cdots \\ \nonumber n_s&=& \frac {\partial f_0}{\partial \mu} +
\frac 1 N \lambda \frac{\partial}{\partial \mu} \langle
E_1(\bm{\tau})\rangle _0 + \cdots \quad .
\end{eqnarray}
Here, we have only given terms up to the first cumulant.
$f_0(\lambda,\mu)$ can easily be calculated from Eq.~(\ref{eq84}).
$\frac {\partial f_0}{\partial \lambda}$ and $\frac {\partial
f}{\partial \mu}$ are \textit{linear} in $\langle \tau _n\rangle
_0$ and $\langle \tau ^2_n\rangle _0$ where:
\begin{equation}\label{eq86}
\langle \tau _n\rangle _0=\frac{A+2B}{1+A+B}\;, \; \langle \tau
_n^2\rangle _0 = \frac{A+4B}{1+A+B}
\end{equation}
with:
\begin{equation}\label{eq87}
A= \exp[-i(\lambda(\tilde{e}_1+\tilde{e}_2)+ \mu),], \; B= \exp
[-i2\lambda(\tilde{e}_1+2\tilde{e}_2)].
\end{equation}
The correction terms in Eq.~(\ref{eq85}), i.e. the first, second,
etc. cumulants are quadratic, cubic, etc. in $\langle \tau
_n\rangle _0$ and $\langle \tau _n^2\rangle _0$. The l.h.s. of
Eq.~(\ref{eq85}) becomes arbitrary small since
$v-v_{gs}\rightarrow 0$ implies $n_s\rightarrow 0$. Then it
follows that $\langle \tau _n\rangle _0$ and $\langle \tau
_n^2\rangle _0$ in leading order are linear in $v-v_{gs}$ and
$n_s$ and that the cumulant terms in Eq.~(\ref{eq85}) are of
higher order. After having performed in Eq.~(\ref{eq85}) the
differentiations, we are allowed to set $\mu(v,\bar{n}_s)=0$,
since we are calculating $\bar{n}_s$ as function of $v$. Writing A
and B as the result for $\epsilon =0$ plus a correction:
\begin{equation}\label{eq88}
A = \bar{n}_s(1 + \delta _A)\;, \; B =
\frac{(v-v_{gs})-(\tilde{e}_1+\tilde{e}_2)\bar{n}_s}{2(\tilde{e}_1+2\tilde{e}_2)}(1+\delta
_B)
\end{equation}
one finds from Eq.~(\ref{eq85}):
\begin{equation}\label{eq89}
\delta_A \simeq (a \bar{n}_s +b(v-v_{gs}))\ln \bar{n}_s, \; \delta
_B \simeq (c\bar{n}_s+d(v-v_{gs}))\ln \bar{n}_s
\end{equation}
in leading order in $v-v_{gs}$ and $\bar{n}_s$. The coefficients
a,b,c and d depend on $\sum \limits _{n_2(\neq 0)}
\tilde{A}_{0n_2}(\epsilon), \sum\limits_{n_2 \neq 0}
\tilde{B}_{0,n_2}(\epsilon), \sum \limits _{n_2\neq 0}
\tilde{C}_{0n_2}(\epsilon)$, etc. and vanish for $\epsilon =0$.
With Eqs.~(\ref{eq88}) and (\ref{eq89}) we can eliminate $\lambda$
from Eq.~(\ref{eq87}) which finally yields:
\begin{equation}\label{eq90}
\bar{n}_s(v) \simeq x^\delta +O(x^{\delta +1}\ln
x)+O(x^{2\delta-1} \ln x)
\end{equation}
with
\begin{equation}\label{eq91}
x= \frac{v-v_{gs} (\epsilon)}{v_-(\epsilon)-v_+(\epsilon)}
\end{equation}
and
\begin{equation}\label{eq92}
\delta (\epsilon) = \frac {v_0(\epsilon)-v_{gs}(\epsilon)}
{v_-(\epsilon)-v_{gs}(\epsilon)}
\end{equation}
where we used Eq.~(\ref{eq82}). Comparison of the result
(\ref{eq90}) with the corresponding one for noninteracting
particles (Eq.~(\ref{eq43})) shows that the interactions do not
change the power law dependence. But, they lead to modified
next-to-leading order corrections, containing logarithmic
dependence on $v-v_{gs}$.

\section{Discussions and conclusions}
\label{5} The purpose of the present paper has been twofold. First
of all, we wanted to establish a relationship between the
stationary points which are solutions of a set of coupled,
\textit{nonlinear} equations, and symbolic sequences
$\bm{\sigma}$, where $\sigma _n$ takes values from an alphabet
${\mathcal{A}}$. We have proven for the class of $\phi^4$-models
that such a unique relation exists, provided that the coupling
parameter $\epsilon$ is below a critical value $\epsilon _c$. This
proof is based on Aubry's anti-continuum limit \cite{17,18} and
the application of the implicit function theorem, similar to the
proof of existence of breather by MacKay and Aubry \cite{19}.
Assuming that such a one-to-one correspondence exists in general
we have demonstrated in the second section that one can use a
``thermodynamic'' formalism for the calculation of statistical
properties of the stationary points. Consequently, this
description allows to investigate topological quantities of the
so-called ``energy landscape'' of a potential energy $V(\vec{x}_
1, \ldots,\vec{x}_n)$. Particularly, it yields an explanation why
a topological singularity, e.g. for the Euler characteristic of
the manifold ${\mathcal{M}}_n(v)$ (cf.~Eq.~(\ref{eq4})) can occur.
We stress that our ``thermodynamic'' formalism is not the same as
that recently used to calculate the full canonical partition
function \cite{22}. There the configuration space has been divided
into basins of the stationary points. Taking also the vibrational
degrees of freedom into account these authors were able to
calculate the canonical partition function, under a couple of
assumptions.

The relationship between stationary points and sequences
$\bm{\sigma}$ already proves that there are in total
$n_{{\mathcal{A}}}^N = \exp (N \ln n_{{\mathcal{A}}})$ stationary
points. $n_{{\mathcal{A}}}$ is the number of ``letters'' of the
``alphabet'' ${\mathcal{A}}$. For rather special models this has
already been shown to be true \cite{15,23,24}. Then it is easy to
calculate the saddle index distribution function $P_N(n_s)$
because the number of negative eigenvalues of the Hessian can be
related to the number of $\sigma_n$'s taking certain values. For
the $\phi^4$-model this value is 0. Simple combinatorics leads for
$P_N(n_s)$ to a Gaussian of width proportional to $N^{-1/2}$ and a
maximum at $n_s^{max}= 1/3$. It is interesting that this value
obtained for a $\phi^4$-lattice model coincides with the result
found for Lennard-Jones clusters for $4 \leq N \leq 9$ \cite{3}.
Is this just an accident? The answer is not clear. It may be true
that the potential energy landscape of a liquid can be decomposed
into basic ``units'' which are double-well-like. Of course, the
smallest ``unit'' one can choose are \textit{two} adjacent local
minima. Because both must be connected by a barrier one arrives at
a double-well potential with extrema being labelled by $+,0,-$.
The main open question is: Can one really construct the full
energy landscape by connecting such double well potentials and
accounting correctly for its connectivity?

For the $\phi^4$-model we have found the energy dependence of the
saddle index. It has become clear that the functions
$n_s(\bar{v})$ and $\bar{n}_s(v)$ are not identical. Whereas both
functions are equal one at the largest possible value for
$\bar{v}$ and $v$ :$\bar{v}_{max}=v_{max}=N^{-1}E(0,\ldots,0)$,
they have different behavior below $N^{-1}E(0,\ldots,0)$.
$n_s(\bar{v})$ vanishes at $\bar{v}^*$ \textit{above} the ground
state energy $v_{gs}$ and $\bar{n}_s(v)$ becomes zero at
$v=v_{gs}$, only. We think that this behavior is true even for
liquid systems. The ``thermodynamic'' formalism has made it
possible to get quantitative results for both functions, at least
for small enough coupling parameter $\epsilon$. $n_s(\bar{v})$ is
linear in $\bar{v}$ for $\epsilon =0$ and becomes nonlinear for
$\epsilon >0$. $\bar{n}_s(v)$ is already nonlinear in v in case of
$\epsilon =0$. For v close to $v_{gs}$ it exhibits a power law. We
have shown that this power law behavior is not changed for
$0<\epsilon < \epsilon_c$, assuming the ground state to be
ferroelastic. For an antiferroelastic ground state one arrives at
the same conclusion. However, whether the power law exists for
periodic ground states with period larger than two or even for
quasiperiodic ones is not yet clear. In addition, its importance
for physically relevant quantities is not obvious.

The fact that functions $n_s$ and $\bar{n}_s$ are different, seems
to be in variance with the numerical results of Ref.~\cite{2}.
These two functions were found to be the same, within statistical
errors. This implies that also $\bar{n}_s(v)$ vanishes at an
energy above $v_{gs}$. The liquid in Ref.~\cite{2} was
equilibrated at a temperature, e.g.~$T=2$ and 0.5 (in
Lennard-Jones units). $T=0.5$ is already close to the mode
coupling temperature $T_c$ where the dynamics is already rather
slow. For $n_s=0$ there is a huge variety of stationary points
with energies $v_{gs}\leq v \leq N^{-1}E(\bm{\sigma})$ with
$\sigma_n=+$ or $-$. Their maximum weight occurs when $N_+=N_-=
N/2$ ($N_\sigma$ = number of $\sigma_n$'s in $\bm{\sigma}$ which
are equal to $\sigma$). The stationary points with $n_s=0$ which
are close to $v_{gs}$ have an exponentially smaller weight and can
only be found at $T$ much smaller than $0.5$ where the
configurations look like a crystal with a low concentration of
defects. It seems to us unlikely that the simulation done in
Ref.~\cite{2} which has reached $T=0.5$ from the {\textit{liquid}}
phase has really been in the range of such defected crystalline
configurations. Our result that $\bar{n}_s(v)$ vanishes at
$v_{gs}$, only, which implies that $\bar{n}_s$ as function of
temperature vanishes at $T=0$, only, is consistent with recent
results (\cite{22} and second paper of \cite{8}).

Let us come back to a liquid system. How could one apply a similar
strategy as used in the present paper? The answer is as follows:
Divide the sample with $mN$ particles into m-boxes with N
particles where $1 \ll m \ll N$. Let us switch off the inter-box
interactions. Then the number of stationary points and the saddle
index properties can be related to the corresponding quantities of
a single box. This starting point corresponds to Aubry's
anti-continuum limit, although a single box represents already a
nontrivial problem. Then one can use again the implicit function
theorem to prove that the topological features are unchanged under
turning on the inter-box interactions, provided their strength is
below a critical value. However, increasing this strength such
that inter- and intra-box- interactions are the same may exceed
the critical strength. As far as we know this type of reasoning
has been used first by Stillinger (see \cite{25} and references
therein). Recently it has been used again \cite{22,26}. Based on
independent boxes the authors of Ref.~\cite{26} have derived a
relationship between $\alpha$, a and $\gamma$ which yield the
number of stationary points with saddle index $n_s=N_s/N$ of a
single box:
\begin{eqnarray}\label{eq93}
M_n(n_s=0)= \exp \alpha N \\ \nonumber M_n (n_s=1/N) = a N \exp
\alpha N
\end{eqnarray}
and the total number of stationary points of a box:
\begin{equation}\label{eq94}
M_N^{tot} = \exp \gamma N.
\end{equation}
This relation is:
\begin{equation}\label{eq95}
\gamma = \alpha + a \ln 2 \;.
\end{equation}
Now let us apply this reasoning to our $\phi^4$-model. There it
is:
\begin{eqnarray}\label{eq96}
M_N(n_s=0)=2^N= \exp(N \ln 2) \\ \nonumber M_N(n_s=1/N)=N \cdot
2^{N-1}= \frac 1 2 N\exp (N \ln 2) \\ \nonumber M_N^{tot}=3^N=
\exp(N \ln 3)\;,
\end{eqnarray}
i.e. we find: $\alpha = \ln 2$, $a = \frac 1 2$, and $\gamma = \ln
3$. Substituting $\alpha $ and a into Eq.~(\ref{eq95}) we get for
its r.h.s.:
\begin{equation}\label{eq97}
\gamma _{r.h.s.}= \frac 3 2 \ln 2 \simeq 1.039
\end{equation}
which is close but not identical to $\gamma = \ln 3 \simeq 1.098$.
Whether this small discrepancy is a hint that correlations between
boxes can be neglected or not, is not obvious.

To conclude, we have demonstrated the usefulness of a one-to-one
correspondence between stationary points and symbolic sequences.

\medskip

{\bf{Acknowledgement}}: I am very grateful for all the discussions
I have had with Serge Aubry. The present paper is a typical
example how his scientific contributions have strongly stimulated
my own research. I also thank D. A. Garanin for his support in
preparing the figures.


\begin{thebibliography}{99}

\bibitem{1}
L. Angelani, R. di Leonardo, G. Ruocco, A. Scala, F. Sciortino,
Phys. Rev. Lett. {\bf{85}} (2000) 5356-5359.

\bibitem{2} K. Broderix, K. K. Bhattacharya, A. Cavagna, A.
Zippelius, I. Giardina, Phys. Rev. Lett. {\bf{85}} (2000)
5360-5363

\bibitem{3} J. P. K. Doye, D. J. Wales, J. Chem. Phys. {\bf{116}}
(2002) 3777-3788

\bibitem{4} R. Franzosi, M. Pettini, Phys. Rev. Lett. {\bf{92}}
(2004) 060601 (1-4)

\bibitem{5} M. Kastner, Phys. Rev. Lett. {\bf{93}} (2004) 150601
(1-4)

\bibitem{6} R. Franzosi, M. Pettini, L. Spinelli, Phys. Rev. Lett.
{\bf{84}} (2000) 2774-2777

\bibitem{7} L. Casetti, E. G. D. Cohen, M. Pettini, Phys. Rev.
E{\bf{65}} (2002) 036112 (1-4)

\bibitem{8} L. Angelani, L. Casetti, M. Pettini, G. Ruocco, F.
Zamponi, Europhys. Lett. {\bf{62}} (2003) 775-781; F. Zamponi, L.
Angelani, L. F. Cugliandolo, J. Kurchan, G. Ruocco, J. Phys.
A{\bf{36}} (2003) 8565-8601

\bibitem{9} A. Andronico, L. Angelani, G. Ruocco, F. Zamponi,
Phys. Rev. E{\bf{70}} (2004) 041101 (1-14)

\bibitem{10} D. A. Garanin, R. Schilling, A. Scala, Phys. Rev.
E{\bf{70}} (2004) 036125 (1-9)

\bibitem{11} D. J. Wales ``Energy Landscapes'', Cambridge
University Press (2003)

\bibitem{12} V. I. Arnold, A. Avez ``Ergodic Problems in Classical
Mechanics'', Benjamin, New York (1968)

\bibitem{13} J. Moser ``Stable and Random Motions in Dynamical
Systems'', Princeton University Press, Princeton (1973)

\bibitem{14} S. Aubry, Solid State Sciences {\bf{8}} (1978)
264-278; in ``Structure et Instabilit\'e'', ed. C. Godr\`eche,
Edition de Physique (1986) 73-194

\bibitem{15} R. Schilling in ``Nonlinear Dynamics in Solids'', ed.
H. Thomas, Springer Verlag (1992) 212-241

\bibitem{16} A. Cavagna, I. Giardino, G. Parisi, Phys. Rev.
B{\bf{57}} (1998) 11251-11257

\bibitem{17} S. Aubry, A. Abramovici, Physica D{\bf{43}} (1990) 199-219

\bibitem{18} S. Aubry in ``Twist Mappings and their
Applications'', eds. R. McGehee, K. R. Meyer, The IMA Volumes in
Mathematics and Applications {\bf{44}}, (1992) 7-54

\bibitem{19} R. S. MacKay, S. Aubry, Nonlinearity {\bf{7}} (1994)
1623-1643

\bibitem{20}
R. S. MacKay, J.-A. Sepulchre, Physica D{\bf{48}} (1995) 243-254

\bibitem{21} S. G. Kratz, H. R. Parks ``The implicit function
theorem'', Birkh\"auser, Boston (2003)

\bibitem{22} M. S. Shell, P. G. Debenedetti, A. Panagiotopoulos,
Phys. Rev. Lett. {\bf{92}} (2004) 035506 (1-4)

\bibitem{23} F. H. Stillinger, T. A. Weber, Phys. Rev. A{\bf{25}} (1982)
978-988

\bibitem{24} P. H\"aner, R. Schilling, Europhys. Lett. {\bf{8}}
(1989) 129-134

\bibitem{25} F. H. Stillinger, Phys. Rev. E{\bf{59}} (1999) 48-51

\bibitem{26} D. J. Wales, J. P. K. Doye, J. Chem. Phys. {\bf{119}},
(2003) 12409-12416



\end{thebibliography}
\end{document}